Packet Error Rate performance of IEEE802.11g under Bluetooth interface

Salim Abukharis[1], Jafar A. Alzubi[2], Omar A. Alzubi[3] , and Saeed Alamri[4]

[1]Elmergib University, Libya, s.abukahris@elmergib.edu.ly

[2]Al-Balqa Applied University, Jordan, j.zubi@bau.edu.jo

[3]Al-Balqa Applied University, Jordan, o.zubi@bau.edu.jo

[4] Albaha University, Saudi Arabia, salamri@bu.edu.sa

**Abstract--**This paper approaches the issues concerning the coexistence of the IEEE 802.11g wireless networks and the ad-hoc Bluetooth networks that operate within the same 2.4 GHz band. The performance of 802.11g is evaluated by simulating the PER (Packet Error Rate) parameter and coverage area. The simulation experiments are based on the worst case scenario presumption, which entails the transmission of HV1 packet which HV1 link requires transmission on 100% of the Bluetooth (BT) time slots using the maximum hop rate of 1600 hops/s. The paper suggests a practical approach to mitigate the interference using symbol erasures technique through the assessment of the PER parameter at various $E_b/N_o$ values. The results show that the symbol erasures can effectively mitigate the interference and enhance the performance of the 802.11g.

Introduction

In 1997 the IEEE 802.11 standardization body for WLANs defined specifications for the Medium Access Control (MAC) sub-layer and three different low-bit rate physical layers (PHY) supporting 1 and 2 Mbps (IEEE 802.11 1999). Due to their limited bit-rate capabilities, the low data rate systems have been used for data traffic only. Two higher speed physical layers were defined in 1999: the 802.11b PHY in the 2.4 GHz ISM band and the 802.11a PHY in the 5 GHz U-NII band. 802.11b offers bit rates up to 11 Mbps while 802.11a offers bit rates of up to 54 Mbps. In 2002, the IEEE Task Group G (802.11g) developed a high-speed extension to the 802.11b (PHY) in the 2.4 GHz ISM band. In this extension, known as IEEE 802.11g, Coded Orthogonal Frequency Division Multiplexing (COFDM) was adopted as the mandatory modulation scheme (IEEE 802.11g 2003).

Bluetooth (BT) technology (Kurose et al 2009) is used in Wireless Personal Area Networks (WPAN) intended for cable replacement and short-distance connectivity. The data rate of BT is equal to 1 Mb/s and a Frequency Hopping Spread Spectrum (FHSS) modulation scheme is used at the PHY. The BT signal hops over the whole ISM frequency band which is partitioned into 79 frequency bins of 1 MHz bandwidth each. In following its hopping pattern, a standard BT device does not respect other signals transmitting in the same band thus leading to coexistence conflicts.

In 802.11g, the OFDM signal occupies 16.5 MHz of the 20 MHz channel bandwidth and up to three non-overlapping channels can be used simultaneously in the ISM band. In contrast, the BT signal hops over 79, 1 MHz wide, channels across the same frequency band. A collision occurs when both the 802.11 and BT packets overlap in both time and frequency. This collision is seen by an 802.11 device as a form of narrow band interference, causing a severe degradation in throughput performance. In order to mitigate this effect, the IEEE 802.15.2 Task Group was created, in order to develop recommendations for coexistence mechanisms, i.e., techniques that would allow 802.11 WLANs and BT to operate simultaneously in the same environment without significantly affecting the performance of one another.

There some studies have been carried out on the performance of 802.11g in the presence of BT interference (Ling et al 2012, Kasem et al 2012, Huang et al 2009). Two classes of coexistence mechanisms have been defined: collaborative and non-collaborative techniques (IEEE 802.15.2 2003). With collaborative techniques it is possible for the BT device and the WLAN to exchange information on mutual operation. Collaborative techniques can be implemented only when the BT and the 802.11 devices are collocated in the same terminal. With non-collaborative techniques there is no way to exchange information between the two networks which operate independently. Examples of collaborative coexistence mechanisms include the META (MAC Enhanced Temporal Algorithm) scheduling scheme (Shellhammer 2004) and the AWMA (alternating wireless medium access) scheme presented in (Liang 2001). Arumugam (Arumugam et al 2003) also investigated the coexistence of 802.11g and the high data-rate Bluetooth and they proposed the following mechanism: since the Bluetooth interference affects only a small number of subcarriers and if the receiver knows where the centre frequency of the BT interferer (through channel estimation or from the BT receiver) then it can inform the Viterbi decoder not to base its decisions on the corrupted subcarriers, i.e., symbol erasure technique.

It can be concluded from the previous discussion that the issue of 802.11b/g and BT interference has received significant levels of attention in both industry and academia. All these studies agreed that when two devices operate in the same environment simultaneously, levels of interference will occur that affects the performance of the 802.11b/g and BT devices. The aim of this paper is to investigate the impact of BT interference on the performance of IEEE 802.11g standard; we also propose the symbol erasures technique to mitigate the effect of Bluetooth interference. Simulation results showed that the simple erasure technique can recover performance to satisfactory levels as long as the OFDM receiver can track the interference without any changes to the 802.11g standard.

## I.  Bluetooth interference model

For a BT transmission to disrupt the packets of an 802.11g transmission there must be an overlap both in time and in frequency, which is illustrated in Figure 1 The likelihood of interference depends on the packet length, the bandwidth occupancy and load factor of both the WLAN and BT systems. The BT interference appears as a narrowband interference to an OFDM signal. The centre frequency of the BT signal hops randomly and uniformly across the 79 channels ranging from 2402 to 2480 MHz. Hence the probability that the BT signal will overlap the OFDM signal in frequency is about 16/79 (i.e. ≈ 0.2 or 20 %) and the BT transmitter is only active for 366 μs in each 625 μs dwell period. The load factor of the BT piconet must also be taken into consideration when determining the overall probability of collision. This interference model is similar to that used in (Wong et al 2003), the time offset, $t_d$, between the start of the 802.11g WLAN transmission and the BT piconet transmission is modelled as a uniformly distributed random variable between 0 and 625 μs. In the simulation program, the interference of the BT signal with the OFDM signal was implemented in the frequency domain. The spectrum of the BT signal is obtained by generating the GFSK (Gaussian Frequency Shift Keying) signal of a BT device in the time domain, then the Fourier transform of the time – domain signal at the sub-carrier frequencies of the OFDM signal is taken. The GFSK signal is generated at passband and an expression for the GFSK passband signal is given in equation (5.1) [68] where $f_{bt}$ is the BT centre frequency. The SIR level determines the BT transmit power $P_{bt}$. The frequency response of the GFSK signal sampled at the sub-carrier frequencies of an OFDM symbol is then obtained by using the *freqz* function in MATLAB.

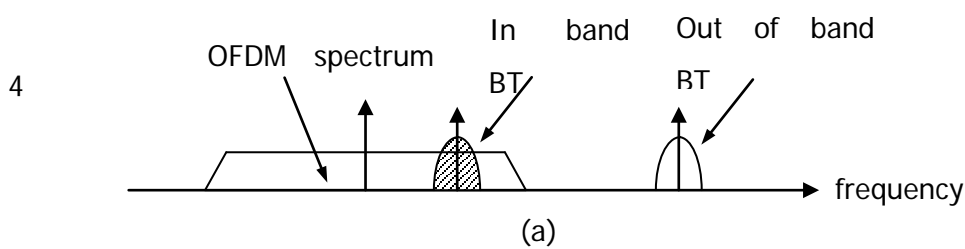

(a)

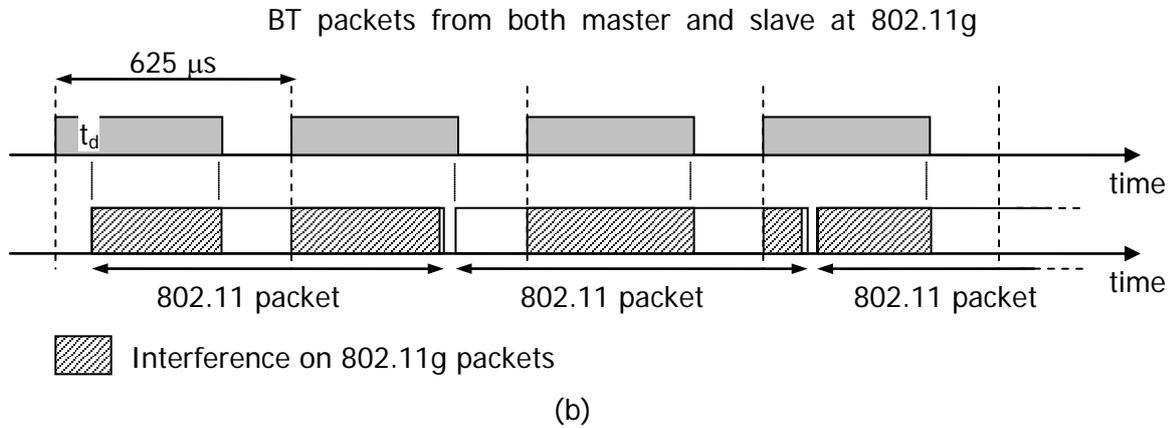

BT packets from both master and slave at 802.11g

Interference on 802.11g packets

(b)

Figure 1. Interference of BT signal on an 802.11g signal due to (a) overlap in frequency, (b) overlap in time

## II. Symbol Erasure Mechanism

The insertion of symbol erasures operates at the PHY layer. The OFDM signal sees the BT signal as a narrow band interference affecting a small number of sub-carriers. The SIR for each independent sub-carrier is determined instantaneously by the power of both the OFDM symbol and the portion of the BT signal transmitted over the bandwidth corresponding to that sub-carrier. Those data symbols corresponding to the sub-carriers with low SIR are replaced by erasures (i.e. setting the received complex modulation symbol to $0+j0$). This approach can be considered as a form of random puncturing of the FEC code. This avoids a large bias to the path metrics in the Viterbi algorithm which would otherwise be introduced by the corrupted data symbols. In addition, the use of bit interleaving across the OFDM symbol helps to reduce the impact of bursts of erasures generated when several adjacent data symbols are corrupted. Since the power of the BT signal mostly is concentrated around its centre frequency, the erasures would be inserted at those OFDM sub-carriers closest to the BT centre frequency. The advantages of using an erasure mechanism are:

1-There is no change to either the 802.11g or BT specifications;

2- Other co-existence mechanisms may also be used;

3-No explicit collaboration between 802.11g and BT is needed.

### III. Results and Discussion

In addition to investigating the $PQM_{rel}$, this sub-section presents results of the simulated PER performance of 802.11g when transmitting 100-byte packets in a multipath channel with $\tau_{rms}$ equal to 100 ns and including BT interference. The legends for the following Figures are: E0, E5 and E7 denote 0, 5 and 7 erasures, respectively. Figure 2 shows curves of PER versus $E_b/N_o$ with the number of symbol erasures as a parameter for 802.11g data rates of 12, 24, 36, 48 and 54 Mb/s in the presence of BT inference. Without erasures, the BT interference substantially corrupts the 802.11g signal, resulting in an error floor at just above PER = 0.1. The insertion of symbol erasures significantly improves performance. For data rates of 12 and 24 Mb/s, 7 erasures improve the PER performance better than 5 erasures, while for data rates of 36 and 54 Mb/s, 5 erasures give a better PER performance. For a data rate of 48 Mb/s, 5 and 7 erasures give similar PER performances. These results demonstrate that the there is a great effect on the IEEE802.11g performance due to Bluetooth interference and symbol erasures can alleviate this effect

In addition to PER measurements, the normalised throughput was used as another metric to evaluate the performance of 802.11g at data rates of 12, 24, 36, 48 and 54 Mb/s when transmitting in the presence of BT interference. The results showed that the number of symbol erasures that required to restore the performance of IEEE802.11g depends on the data rate of 802.11g.

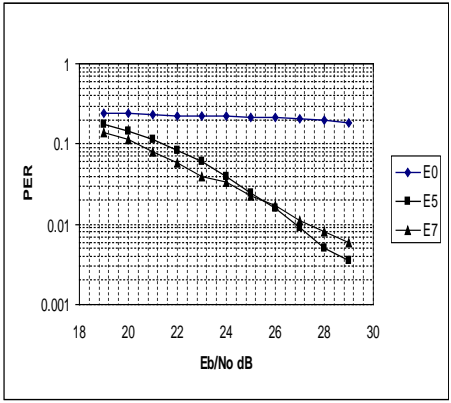

(a) 12 Mb/s

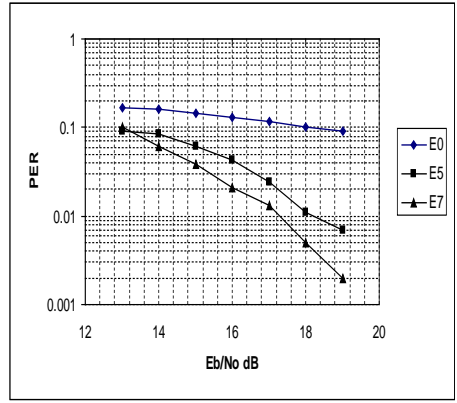

(b) 24 Mb/s

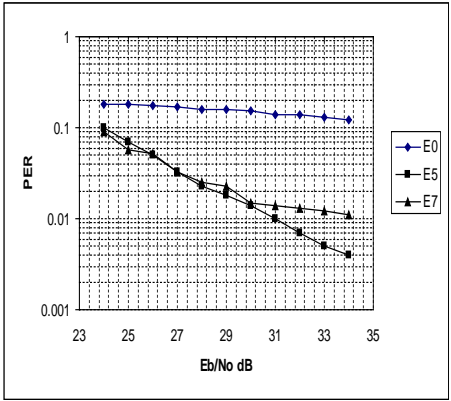

(c) 36 Mb/s

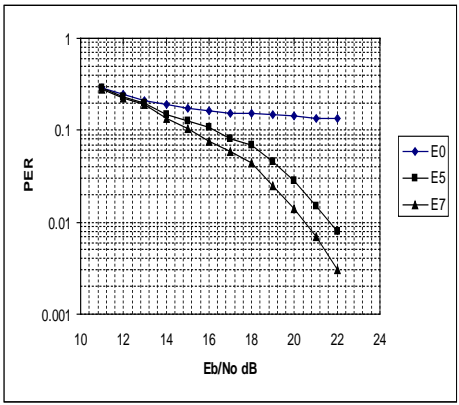

(d) 48 Mb/s

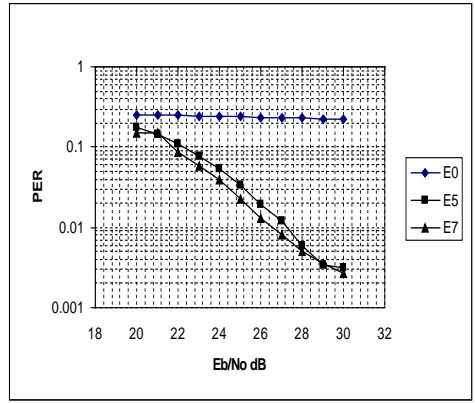

(e) 54 Mb/s

Figure 2. PER versus Eb/No with number of erasures as a parameter for 802.11g in the presence of BT interference at data rates: (a) 12, (b) 24, (c) 36, (d) 48 and (e) 54 Mb/s

## IV. Conclusions

In this paper we investigated the system performance of 802.11g when 100 byte packet over an IEEE 802.11g PHY in the presence of BT interference. Both the conventional throughput and the new perceived video quality metrics were used to characterize the 802.11g coverage in the presence of BT interference when MPEG-2 encoded video clips are transmitted. The results showed that video quality in 802.11g WLANs is substantially degraded by BT interference. The symbol erasure technique was used effectively to restore the video quality. However, there is a trade-off between the number of erasures used, the 802.11g data rates and the encoded video data rates. The relative perceived video quality metric (i.e. $PQM_{rel}$) was used to find the required number of erasures in an attempt to restore the video quality before allowing 802.11g to change to a more robust modulation format and lower data rate. The required number of erasures that were found from these investigations are between 5 and 7. The effect of the Bluetooth inference on video transmission over IEEE802.11g is a subject of on-going research by the authors.